\documentclass[useAMS,usenatbib]{mn2e}

\usepackage{epsfig}
\usepackage{amssymb}
\usepackage{natbib}
\usepackage{aas_macros}

\def \apj {ApJ}
\def \apjs {ApJS}
\def \apjl {ApJL}
\def \aap {A\&A}

\def\Lcorr{$L_r^{\rm corr}$}

\def\re{$R_e$}

\def\msun{${\rm M_{\odot}}$}
\def\lsun{${\rm L_{\odot}}$}
\def\ML{$M_{\rm dyn}/L_r$}
\def\MsL{$M_{\rm star}/L_r$}
\def\sisL{$\sigma^2/L_r$}
\def\ReL{$R_e/L_r$}

\def\mstar{$M_{\rm star}$}

\def\sis{$\sigma$}

\def\ls{\lower 2pt \hbox{$\;\scriptscriptstyle \buildrel<\over\sim\;$}}
\def\gs{\lower 2pt \hbox{$\;\scriptscriptstyle \buildrel>\over\sim\;$}}
\def\mbh{$M_{\rm BH}$}
\def\mstar{$M_{\rm star}$}

\begin{document}

\title[Ages in the size-mass relation]
{The age dependence of the size-stellar mass relation and some implications}

\author[Shankar \& Bernardi]
{Francesco Shankar$^{1,2}$\thanks{E-mail:$\;$shankar@mpa-garching.mpg.de}, Mariangela Bernardi$^{3}$\thanks{E-mail:$\;$bernardm@physics.upenn.edu}\\
$1$ Max-Planck-Instit\"{u}t f\"{u}r Astrophysik,
Karl-Schwarzschild-Str. 1, D-85748, Garching, Germany\\
$2$ Department of Astronomy,
The Ohio State University, Columbus, OH 43210\\
$3$ Department of Physics and Astronomy
University of Pennsylvania, 209 South 33rd St,
Philadelphia, PA 19104}

\date{}
\pagerange{\pageref{firstpage}--\pageref{lastpage}} \pubyear{2009}
\maketitle

\label{firstpage}


\begin{abstract}
We use a sample of about 48,000 SDSS early-type galaxies to show
that older galaxies have smaller half-light radii \re\ and larger
velocity dispersions \sis\ than younger ones of the same stellar
mass \mstar. We use the age-corrected luminosity \Lcorr\ as a proxy
for \mstar\ to minimize biases: below \Lcorr$\sim 10^{11}$\lsun,
galaxies with age $\sim 11$~Gyrs have \re\ smaller by 40\% and \sis\
larger by 25\%, compared to galaxies that are 4~Gyrs younger. The
sizes and velocity dispersions of more luminous galaxies vary by
less than $15\%$, whatever their age, a challenge for
current galaxy formation models.
A closer check reveals that the lowering in the dispersion
is caused by older galaxies that show a significant
departure from the
\re--\Lcorr\ and \sis--\Lcorr\ relations at high \Lcorr. Such features
might find an explanation in models where more massive galaxies
undergo more minor mergers than less massive galaxies at late times,
thus causing a break in the homology. In
terms of the Fundamental Plane of early-type galaxies, the data
indicate that all galaxies
show a significant and similar increase in the dynamical-to-stellar mass ratio
with increasing mass, independent of their age. However, older galaxies
have smaller $M_{\rm dyn}/$\mstar\ ratios than objects which
formed more recently. These findings may suggest that lower mass galaxies
and, at fixed stellar mass, higher redshift galaxies,
formed
from gas-richer progenitors, thus underwent
more dissipation and contraction.
\end{abstract}


\section{Introduction}\label{sec|intro}
The formation and evolution of galaxies is still hotly debated. The
cooling of baryons in dark matter halos should form compact and
dense self-supporting, rotating stellar and gaseous disks \citep[e.g.,][]{Fall80,Navarro00,Governato07}. Later major mergers between disk galaxies have then been
proposed as the main routes to form elliptical galaxies \citep[e.g.,][]{Toomre}. Several detailed numerical simulations
\citep[e.g.,][]{Barnes91,Boylan06,Dekel06,Robertson06,Burkert08,Hop08FP} have shown that many dynamical and photometrical
properties of the remnant spheroidal galaxies can be explained
simply in terms of the merging of progenitors having varying
levels of gas-richness. Galaxies which form from gas-rich,
dissipative, mergers result in more compact remnants with larger
velocity dispersions.

On the other hand, in a pure monolithic model of galaxy formation
\citep[e.g.,][]{Eggen62}, stars are formed in a single burst of
star formation from gas falling towards the center, and the
evolution is passive thereafter. Although there is clear evidence
for a red and dead population of massive early-type galaxies (see
\citealt{Renzini06} for a review), hierarchical merging could still have played
some role at late times.  The metallicities of typical early-type
galaxies are well reproduced in models with frequent minor mergers
at moderate redshifts \citep[e.g.,][]{Bournaud07,Naab07} and are not much
affected by later dry mergers \citep{Pipino08}.  The sizes and velocity dispersions of
BCGs in the local Universe are evolving in a manner which suggests
frequent minor dry mergers as recently as 1 Gyr ago \citep{Bernardi09}.
The clustering
and number density of massive galaxies in the Sloan Digital Sky
Survey (SDSS), the 2dF-SDSS LRG and QSO Survey (2SLAQ), the NOAO
Deep Wide-Field Survey and in DEEP2 also suggest that some merging
events involving massive galaxies must have occurred since redshift
$z\sim 1$ \citep[e.g.,][]{Bundy07,White07,Wake08}, but that the majority of the stellar mass had
already been assembled by this time.

In addition, there is now growing evidence that massive galaxies at
$z\sim 2$ are much smaller and denser than their local counterparts
of the same stellar mass \citep[e.g.,][]{Trujillo06,VanDokkum08,Cimatti08,Saracco08}. These
observations are in line with the idea that high--redshift galaxies
formed in a denser universe, and therefore from baryonic clumps
which collapsed in denser, more gas-rich environments, which, in
turn, induced more dissipation, more compact remnants and higher
velocity dispersions.
However, similarly compact galaxies to those observed at high-redshift
do not exist in the local universe, raising the question of what process
or processes have acted to increase the sizes of these objects to
make them consistent with the larger sizes we see at late times
\citep[e.g.,][]{VanDokkum08}.


\begin{figure*}
\includegraphics[width=17.5truecm]{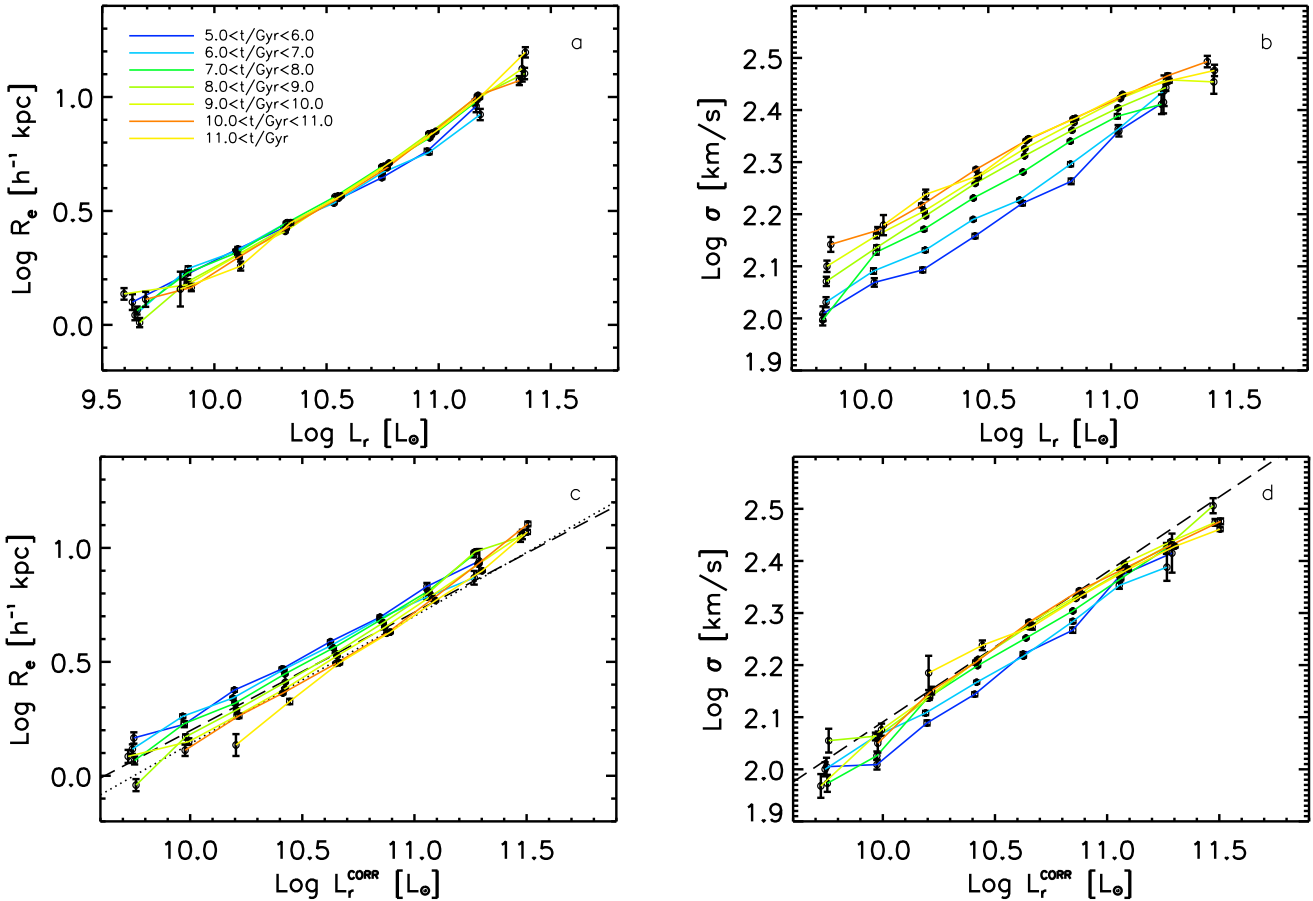}
\caption{\emph{Top}: size \re\ (\emph{left}) and velocity dispersion
 \sis\ (\emph{right}) as a function of $r$-band luminosity $L_r$,
 for galaxies of different ages, as labeled.
 \emph{Bottom}: Same as top panels, but after correcting luminosity
 for age effects, so that it is a proxy for stellar mass. The \emph{long-dashed} lines in Figs.~1c and 1d are the fits to the \re-\Lcorr\ and \sis-\Lcorr\ correlations to the subsample of galaxies with age $t<6$ Gyr, displaced in normalization to match the locus defined by older galaxies. The \emph{dotted} line in
 Fig.~1c, with arbitrary normalization, has a slope of $0.56$ as derived by \citet{Shen03} fitting the \re-\Lcorr\ relation to the whole sample of SDSS early-type galaxies.}
 \label{fig|RV}
\end{figure*}

In this Letter, we present evidence that the sizes of early-type
galaxies are difficult to reconcile with a pure monolithic model, at
least for the most massive galaxies. In \S~\ref{sec|data} we
describe the data set and present the measurements on which our
conclusions are based.
We discuss our results and their implications in
\S~\ref{sec|discuconclu}.  Where necessary, we set the cosmological
parameters $\Omega_m=0.30$, $\Omega_\Lambda=0.70$, and
$h\equiv H_0/100\, {\rm km\, s^{-1}\, Mpc^{-1}}=0.7$.

\section{Data}
\label{sec|data}

We use the SDSS-based sample of early-type galaxies from \citet{Hyde08a} who give a prescription for how to correct the SDSS
photometric parameters for known sky subtraction problems which
affect objects with large apparent brightnesses. The sample, which
contains about 48,000 early-type galaxies, is distributed within
the redshift range $0.013 < z < 0.3$, which corresponds to a maximum
lookback time of 3.5 Gyr. The galaxies in the sample have apparent
magnitudes $14.5\lesssim m_r \lesssim 17.5$ (based on deVaucouleur
fits to the surface brightness profiles), axis ratios $b/a>0.6$, and
ages greater than 2 Gyr.  To study the bulk of the early-type
population of local galaxies, we remove the Brightest Cluster
Galaxies (BCGs) from our sample, as they might have had unusual
formation histories (see \citealt{Bernardi09}).  This was done by matching
the \citet{Hyde08a} sample to the maxBCG \citep{Koester07}
and C4 \citep{Miller05} cluster samples (this procedure should
remove most of the BCGs although some contamination could still be
present). Our results do not vary significantly if BCGs were not
excluded. The luminosities we report are estimated by first
correcting the absolute magnitudes for evolution to $z=0$ (following
\citealt{Hyde08a}, we add $0.9z$ to the observed magnitudes),
and then setting $\log(L_r/L_\odot) = -0.4(M_r-4.62)$ \citep{Blanton01}. Estimated stellar masses and ages for these objects are
from \citet{Gallazzi05}.

\section{Results}
\label{sec|results}

We divide the sample into bins of different ages, as labeled in
the top left panel of Fig.~\ref{fig|RV}, chosen so as to provide at
least 2,000 galaxies in each bin. For each age bin, we further
divide the sample into nine equally-spaced bins in luminosity, and
plot only bins with more than 5 galaxies. We compute the median
size, velocity dispersion, and mass-to-light ratio in each bin of
redshift and luminosity. Uncertainties on these median quantities
are approximated by the square root of the variance divided by the
square root of the number of sources in the bin. Finally, we only
consider galaxies older than 5 Gyrs in our final sample, in order
to minimize contamination from objects for which the luminosity-
and mass-weighted ages might differ substantially.

The top panels in Figure~\ref{fig|RV} show the \re$-L_r$ and
\sis$-L_r$ relations for galaxies of different ages. For
$L_r<10^{11}L_\odot$, lines of constant age run parallel to the
\sis$-L_r$ relation, with older galaxies being offset to larger
\sis.  This is remarkably consistent with previous expectations
which were based on a {\em very} different analysis \citep{ForbesPonman,Bernardi05}. Even more remarkable is the
fact that the \re$-L_r$ relation is almost independent of age. The
age estimate is noisy, so one might have worried that this has
erased any age-dependence in the \re$-L_r$ relation. However, this is
unlikely to be the case, since the \sis$-L_r$ relation does vary for galaxies
of different age.

Because the luminosity changes as the stellar population ages
(typically as $L_r\propto t^{-0.75}$), we would have liked to
replace $L_r$ with the stellar mass \mstar, and study the
\re--\mstar\ and \sis--\mstar\ relations instead.  However, such
plots are complicated by the fact that the SDSS is magnitude
limited and because the age and \mstar\ estimates are highly correlated.
The flux limit will tend to select brighter sources at fixed \mstar, thus
scattering sources
to lower \mstar$/L$ ratios and to younger ages with respect to their
true ones, given the intrinsic correlations in the errors. Therefore,
the oldest objects corresponding to a given \mstar\ may be missing. However, the correct
spread in ages at fixed $L$ is reproduced if luminosities are taken
as independent variables
(we refer the reader to the full analysis presented in Appendix~A of \citealt{Bernardi09} for more discussion).
Therefore, we correct each luminosity for its fading with age by
setting
\begin{equation}
 \log L_r^{\rm corr}=\log L_r+0.75\log (t/5.5~{\rm Gyrs}),
 \label{eq|Lcorr}
\end{equation}
where $t$ is the age.  \citet{Bernardi09} shows that
\Lcorr\ defined in this way is a good proxy for \mstar.

Fig. 1c shows the \re-\Lcorr\ relation for galaxies of different ages. To produce
Figs. 1c and 1d we
first rebin the whole sample in the new grid of luminosities \Lcorr\
and then recompute the correlations, although
we also note that a simple median rescaling of the curves in Fig.~1a and 1b via Eq.~(\ref{eq|Lcorr}) would yield similar results.
In contrast to the \re-$L_r$ relation shown in Fig. 1a, there is now a
clear trend with age:  at fixed \Lcorr$\propto$\mstar,
younger galaxies tend to have larger sizes.
At $\log$(\Lcorr/\lsun) $\sim 10.5$, the offset is $\sim 0.15$ dex;
it decreases to $\lesssim 0.1$~dex at higher \Lcorr.
Fig. 1d shows instead that the $\sigma-$\Lcorr relation is
less age dependent than the $\sigma-L_r$ relation in Fig. 1b.
At log(\Lcorr/\lsun) $\sim 10.5$, the spread in velocity dispersions
is $< 0.1$~dex, and it decreases to $\lesssim 0.05$ dex at
larger \Lcorr.  Above log(\Lcorr/\lsun) $\sim 11$, the relation is
curved -- older galaxies have a lower \sis\ than expected from
extrapolating the \sis-\Lcorr\ relation defined at lower
luminosities to higher \Lcorr. More specifically, we find that
a fit to the sample of young galaxies with age $t<6$ Gyr yields
a slope in the \re-\Lcorr\ relation of $\sim 0.52$ (long-dashed line in Fig.~1c), very close
to the slope of $0.56$ derived by \citet{Shen03} fitting the \re-\Lcorr\ relation to the whole sample of SDSS early-type galaxies (dotted line). (Both the dashed and dotted lines in Figs.~1c and 1d are displaced in normalization to match the locus defined by older galaxies.) Fig.~1c shows that very old ($t>9$ Gyr) galaxies tend to follow an \re-\Lcorr\ relation
with a very similar slope, although more massive ($\log$(\Lcorr/\lsun) $\gtrsim 11$) systems show a systematic and significant deviation, of up to $\Delta \log R_e\sim 0.1$ dex, from the extrapolation
of such a straight line. Analogously, Fig. 1c shows that the same subsample of old and massive galaxies, also show a significant departure of $\Delta \log \sigma\sim 0.05$ dex above $\log$(\Lcorr/\lsun) $\gtrsim 11$ from the straight,
long-dashed line of slope $\sim 0.29$, derived from fitting the \sis-\Lcorr\ relation
calibrated on younger galaxies only. In the overall, this curvature is similar to that
found for BCGs, for which it is even more evident \citet{Bernardi09}. We stress here that this gradual steepening and corresponding flattening in the relations, are
present in galaxies of a fixed age, thus mirroring a clear break in the
homology when moving from lower to more massive systems.

\section{Discussion and Conclusions}
\label{sec|discuconclu}



In the simplest galaxy evolution models, the age of the stellar
population reflects the time of assembly of the galaxy. Hence, older
galaxies are expected to have smaller sizes and larger velocity
dispersions than their younger counterparts of the same \mstar, both
because the high-redshift Universe was denser, and because the
objects at that time are thought to have formed from gas-richer
progenitors. However, the differences observed between the sizes of
old and young galaxies in our sample are far less than what expected
given the evolution \re$\propto (1+z)^{-1}$ at fixed stellar mass
which would result if the galaxy density is proportional to the
density of the universe. For example, galaxies as old as 12 Gyr ($z\sim 4$), should
be displaced by $\Delta \log R_e \sim 0.5$ (i.e., a factor of $\sim 3$) downwards
with respect to the younger galaxies in our sample.
So the absence, in our data, of a strong age-dependent trend with
size, rules these models out.


A more elaborate model postulates that although the sizes were
initially smaller (so as to be consistent with the $z\sim 2$
observations mentioned in the Introduction), they have since
evolved, while the stellar mass has remained unchanged
\citep{Fan08}. This model exploits the fact that the epoch
when early-type galaxies were forming stars is close to that when
AGNs were most active \citep[e.g.,][]{CattaneoBernardi,Granato04,Haiman07,ShankarMsigma,SWM}.
So, in this model, AGN activity is assumed to expel gas from the
central regions; the sudden reduction of mass in the core makes
the surrounding stellar distribution puff up, increasing the size.
Because the objects are assumed to eventually settle back into
virial equilibrium at these larger sizes, with no change in mass,
this model also predicts that the velocity dispersions decrease
from their initial values, but that the age-dependence in the
\sis-\mstar\ relation is not erased.

The Fan et al. (2008) model was calibrated to reconcile the differences
between the $z\sim 2$ and local \re-\mstar\ relations.  At lower
\mstar, it predicts that younger galaxies should be larger by
$\sim 0.15-0.2$~dex, in good agreement with our Fig. 1c.
At higher masses, we find an offset of $\lesssim 0.1$ dex,
which is less than the $\sim 0.3$~dex they predict.
They also predict that
\sis\ should be larger for older galaxies:  they find an increase
of $\sim 0.15$ dex in \sis\ when moving from younger to older galaxies
at large masses, with slightly smaller trends at lower masses.
While we are in qualitative agreement with their predictions, our
results point to a much smaller offset, especially at
$\log$(\Lcorr$/$\lsun)$\, \gtrsim 11$. However, given the
systematic uncertainties in computing the profile and luminosity-dependent
normalization coefficients
in the virial relations (see Fan et al. 2008 for details), it is
difficult to make detailed comparisons with their predictions.
Only ad-hoc numerical simulations
will be able to further probe their model.

In hierarchical models, the stellar population can be older than the
time at which the total mass was assembled into one system \citep[e.g.,][]
{Bower06,DeLucia06}. This is accomplished by
making dense progenitors through wet (gas rich) mergers at high
redshift, followed by a sequence of dry, dissipationless mergers
which serve to reduce the densities \citep[e.g.,][]{Kormendy08}.
However, this evolution in sizes cannot be directly tested from Fig.
1c, which shows galaxies only in their present form. On the other
hand, the observed break in the \sis--\Lcorr\ relation (Fig. 1d),
when combined with the steepening in the \re--\Lcorr\ relation (Fig.
1c), might be a signature that minor mergers played a role in
the mass assembly of at least the most massive objects.

Preliminary results from Shankar et al. (2009b, and references
therein) show that major dry mergers are uncommon for intermediate
mass galaxies, with even the most massive galaxies experiencing one
such event at most. Early-type galaxies, born at $z\sim 2$ and with
$z=0$ mass \mstar$\gtrsim 3\times 10^{10}$\msun, undergo at least
$3-7$ minor dry mergers, with the number increasing with $M_{\rm
star}$. As sketched by Bernardi (2009), and theoretically
discussed by, e.g., \citet[][and references therein]{Ciotti08},
repeated minor mergers of mass ratio $f<1$
can enable the remnants to increase their masses by a factor
$(1+f)$, the sizes by a factor of $(1+2f)$, and to decrease
$\sigma^2$ by a factor of $(1-f)$, thus without changing the virial product $\sigma^2R_e$
much. For example, even 5 minor mergers with $f$ as low as $f\approx 0.2$, would
be capable of increasing the sizes by a factor of $\sim 5$, and the stellar mass
by just a factor of 2.5. Ad-hoc, recent numerical simulations are showing
that this can actually be possible \citet[e.g.,][and references therein]{Naab09}.
This steep and fast evolution in the \re-\mstar\ plane
is another way to efficiently puff-up the high-$z$, compact galaxies.
The main challenge for hierarchical models would then be to grow \emph{all} galaxies of
different ages coherently on a similar size-luminosity
relation, as we see it today (see Shankar et al. 2009b). Also,
hierarchical models tend to produce too large size
dispersions at fixed galaxy luminosity, at variance with our data \citep[][but see also Khochfar \& Silk 2006]{GonzalezSAM08}.
Nevertheless, a late evolution driven by minor mergers in the most massive and older galaxies, that preferentially sit in richer environments,
might explain the gradual steepening of the size-mass relation
at high \Lcorr\ and the corresponding curvature at high \sis. 

\begin{figure}
\includegraphics[width=8.5truecm]{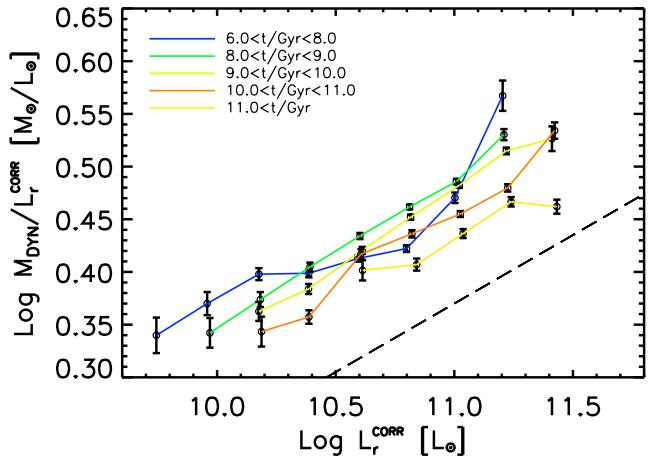}
\caption{Dynamical mass-to-light ratio as a function of age-corrected
 luminosity for galaxies of different ages, as labeled. 
 Irrespective of age, we find that
 all galaxies show a significant tilt.}
\label{fig|tiltTotal}
\end{figure}

Before concluding, we would like to discuss our findings in the
context of the Fundamental Plane relation
 $R_e\propto \sigma^a I^b$
(FP, e.g. \citealt{DDFP}. Observations suggest that
$a\sim 1.43 \pm 0.05$ and $b=-0.79\pm 0.02$ with small scatter
\citep[e.g.,][]{Hyde08b}.  The fact that $(a,b)\ne (2,-1)$ is
sometimes called the `tilt', and is thought to reflect the fact that
\ML\ or \MsL\ is not constant \citep[e.g.,][]{DOnofrio06,Hyde08b}. The idea is that if
\begin{equation}
 \sigma^2\propto \frac{M_{\rm dyn}}{R_e}
    \propto \left(\frac{M_{\rm dyn}}{L_r}\right)\,I_r\, R_e
    \propto L_r^\gamma\,I_r\, R_e ,
 \label{FP2}
\end{equation}
then
\begin{equation}
 R_e\propto \sigma^{2/(2\gamma + 1)} I_r^{-(\gamma+1)/(2\gamma+1)},
 \label{ReFP}
\end{equation}
with $I_r$ the surface brightness of the galaxy.  Previous work has
shown that $\gamma>0$ if $L_r$ is the optical luminosity.
In the discussion which follows, we consider the effect of replacing
$L_r$ with \Lcorr.  Figure~A5 in Bernardi (2009) shows that
\mstar$/$\Lcorr\ vs \Lcorr\ is flat for these galaxies, so this
should be equivalent to studying the stellar mass Fundamental Plane
\citep[e.g.,][]{Hyde08b}.

Fig. 2 shows $M_{\rm dyn}/L_r^{\rm corr}$ versus $L_r^{\rm corr}$,
for galaxies of different bins in formation time.  (We define the
dynamical mass as
 $(M_{\rm dyn}/M_\odot) = 10^{10}\,(\sigma/200{\,\, \rm km\,s^{-1}})^2
(R_e/{\rm h^{-1}\, kpc})$, and only show bins in which there were more than
100 galaxies.)  It is clear that the relation is not flat: except
for the bin with the most recent formation time (which may be
contaminated by selection effects and/or errors in age of the type
discussed by Bernardi 2009), the `tilt' is $\gamma \approx 0.13$ (long-dashed line
in Fig. 2),
and it is approximately \emph{independent} of age.
Under the reasonable assumption of a universal DM profile
\citep[e.g.,][]{NFW}, the tilts reported in Fig.~2 suggest that less massive galaxies are
more concentrated, possibly due to more dissipation, thus inducing
more contraction and a lower dark matter fraction within \re. Note that, as discussed in \S~\ref{sec|data},
all \re\ in SDSS are calibrated
with deVaucouleur
fits (i.e., with a S\'{e}rsic index $n=4$), thus
the non-homology seen in Fig.~2 should principally
derive from actual dynamical mass variations
with stellar mass and not, for example, from non-homology effects
in the light distributions
\citep[e.g.,][and references therein]{Tortora09}.
Moreover, Fig.~2 also shows that, at fixed \Lcorr, galaxies which formed earlier
tend to have smaller $M_{\rm dyn}/L_r^{\rm corr}$ than younger ones.
This offset is mainly driven
by the smaller sizes \re\ associated to older galaxies, as expected if the latter
were formed in a denser and gas-richer environment.

Finally, note that setting $\gamma = 1/4$ in equation~(\ref{ReFP})
would make the Fundamental Plane relation
 $R_e\propto \sigma^{1.33}\,I_{\rm corr}^{-0.83}$
for populations of a fixed formation time.
However, because of the dependence of the zero-point of the
$M_{\rm dyn}/L_r^{\rm corr} -$ \Lcorr relation on formation
time, the slope $\gamma$ becomes shallower if one averages over
a range of formation times.  A smaller value of the tilt $\gamma$
means that the FP coefficient $a$ should be larger when one averages
over the full early-type population than when one restricts the study
to a small range of formation times.  All the galaxies in a cluster
tend to have similar formation times (e.g. Bernardi 2009).
This suggests that the FP computed for a single cluster should have
greater `tilt' (the coefficient $a$ should be further from 2) than
the FP for the full population.  So it is interesting that $a\approx 1.6$
for the full population \citep{Hyde08b}.  Perhaps this is
why the traditional FP, with $I_r$ instead of $I_{\rm corr}$,
has $a\sim 1.43\pm 0.05$ for SDSS early-types \citep{Bernardi03,Hyde08b}, whereas $a\sim 1.24\pm 0.07$ for the Coma cluster
is smaller \citep[e.g.,][]{Jorgensen96}.

\section*{Acknowledgments}
FS acknowledges partial support from NASA Grant NNG05GH77G. MB is
grateful for support provided by NASA grant LTSA-NNG06GC19G. We also
thank Alessandro Bressan, Luigi Danese, Gianluigi Granato, Guinevere
Kauffmann, Andrea Lapi, Federico Marulli, Ravi Sheth, David
Weinberg, and Simon White, for helpful and interesting
conversations.

\bibliographystyle{mn2e}
\bibliography{../../RefMajor}

\label{lastpage}

\end{document}